\documentclass[aps,pre,print,twocolumn,tightenlines]{revtex4}
\usepackage{amsfonts}
\usepackage{amsmath}
\usepackage{amssymb}
\usepackage{graphicx}
\usepackage{url}
\usepackage{hyperref}
\begin{document}

\title{Metastable soliton necklaces supported by fractional diffraction and
competing nonlinearities}
\author{Pengfei Li$^{1,2}$} 
\email{lpf281888@gmail.com}
\author{Boris A. Malomed$^{3,4}$}
\author{Dumitru Mihalache$^{5}$}

\affiliation{$^{1}$ Department of Physics, Taiyuan Normal University, Jinzhong, 030619, China}
\affiliation{$^{2}$ Institute of Computational and Applied Physics, Taiyuan Normal University, Jinzhong, 030619, China}
\affiliation{$^{3}$ Department of Physical Electronics, School of Electrical Engineering, Faculty of Engineering, and Center for Light-Matter Interaction, Tel Aviv University, Tel Aviv 69978, Israel}
\affiliation{$^{4}$ Instituto de Alta Investigaci\'{o}n, Universidad de Tarapac\'{a}, Casilla 7D, Arica, Chile}
\affiliation{$^{5}$ Horia Hulubei National Institute of Physics and Nuclear Engineering, Magurele, Bucharest, RO-077125, Romania}

\begin{abstract}
We demonstrate that fractional cubic-quintic nonlinear Schr\"{o}dinger
equation, characterized by its L\'{e}vy index, maintains ring-shaped soliton
clusters (``necklaces") carrying orbital
angular momentum. They can be built, in the respective optical setting, as
circular chains of fundamental solitons linked by a vortical phase field. We
predict semi-analytically that the metastable necklace-shaped clusters
persist, corresponding to a local minimum of an effective potential of
interaction between adjacent solitons in the cluster. Systematic simulations
corroborate that the clusters stay robust over extremely large propagation
distances, even in the presence of strong random perturbations.
\end{abstract}

\maketitle


\section{Introduction}

\label{Sec I}

Ring-shaped soliton clusters, constructed as coherent chains of fundamental
solitons, are quasistationary multisoliton bound states. These complexes
generalize the concept of necklace-ring beams, which were first studied in
self-focusing Kerr media \cite%
{Soljacic-Segev1,Soljacic-Segev2,Soljacic-Segev3}, and later extended to the
nonlinear Schr\"{o}dinger equation (NLSE) with saturable nonlinearity \cite%
{Desyatnikov-Kivshar1}.

Usually, many-soliton circular structures tend to destroy themselves through
expansion or collapse, an experimentally relevant issue being how to
maintain their quasi-stationary propagation. To balance the destructive
trends, a vortical phase profile can be imposed onto the clusters \cite%
{Desyatnikov-Kivshar2}. In this context, various nonlinearities, including
saturable \cite{Desyatnikov-Kivshar3}, competing quadratic-cubic \cite%
{Kartashov-Cra-Mih-Torner}, cubic-quintic \cite{Mihalache-PRE68-046612}, and
nonlocal \cite{Buccoliero-PRL-98-053901}, were employed to enhance the
robustness of the circular soliton clusters. It was also demonstrated that
two-dimensional (2D) soliton clusters can be supported in active systems,
such as pumped nonlinear optical cavities \cite%
{Vladimirov-Skryabin-PRE65-046606,Skryabin-Vladimirov-PRL89-044101}. A
possibility of the existence of 3D soliton clusters, composed of
spatiotemporal optical solitons, has also been explored \cite%
{Perez-Garcia-Vekslerchik,Crasovan-PRE67-046610,Mihalache-JOB-6-S333}.
Creation of necklace beams was experimentally demonstrated in
photorefractive \cite{Desyatnikov-JOSAB-19-586}, nonlocal \cite%
{Rotschild-OL-31-3312}, and Kerr \cite{Grow-PRL99-133902} media. More
recently, assemblage of metastable necklace clusters of 2D
\textquotedblleft quantum droplets\textquotedblright\ was predicted \cite%
{Kartashov-Malomed-Torner} in the framework of the Gross-Pitaevskii
equations with the cubic nonlinearity modified by the logarithmic factor,
which is induced by the Lee-Huang-Yang correction to the mean-field dynamics
of Bose-Einstein condensates (BEC) \cite{AstraPetrov}. Further exploration
of physically relevant settings which maintain necklace complexes in a
quasi-stable form remains a relevant objective for theoretical and
experimental work. In particular, the creation of nearly stable necklace
patterns with imprinted overall vorticity in a medium with the cubic-quintic
(CQ) nonlinearity (liquid carbon disulfide) was very recently reported in
Ref. \cite{Albert}.

In this connection, it is relevant to stress that the vortical soliton
clusters and usual axisymmetric vortex solitons \cite%
{Michinel,Pego,PhysD,LPF-Mal-Mih2} are different types of beams carrying the
orbital angular momentum (OAM). The clusters carry OAM in the explicit form,
being rotating azimuthally modulated necklace-like structures, while the
vortex solitons carry OAM in the \textquotedblleft hidden\textquotedblright\
form, being axisymmetric structures, while OAM is stored in their phase
circulation. Circular soliton clusters, considered in the present work, are
constructed as quasi-steady states, realizing local minima of an effective
potential. Vortex solitons are fully stationary modes with a topological
charge, \emph{s}, representing the vorticity. Accordingly, stable vortex solitons
keep their shape in the course of the propagation, while unstable ones are
broken up by azimuthal perturbations in several fragments, whose number is
usually (\emph{s}+1). On the other hand, the circular clusters
(\textquotedblleft necklaces\textquotedblright\ ) should be created differently, as they do not
emerge as result of the splitting of unstable vortex solitons. In
particular, the number of separate elements
(\textquotedblleft beads\textquotedblright\ ) in the necklace is independent of \emph{s}. Unlike the
uniform vortex modes with the \textquotedblleft hidden\textquotedblright\
vorticity, for clusters their steadiness is a major issue.

Theoretical studies of optical solitons in NLSE models were lately extended
in two noteworthy directions. On the one hand, the theory has been expanded
for non-Hermitian optical systems, which are modeled by NLSEs with
parity-time-symmetric complex potentials, see reviews. \cite%
{RMP-PT-88-035002,Suchkov,Feng,El-Ganainy} and an experimental report \cite%
{Peschel}. On the other hand, much interest has been drawn to the fractional
Schr\"{o}dinger equations (FSEs), which were originally proposed by Laskin
\cite{Laskin1,Laskin2,Laskin3} and, in a rigorous mathematical form, by Hu
and Kallianpur \cite{Hu}. In the quantum theory, FSE was derived \cite%
{Laskin1} as a model in which the consideration of Feynman path integrals
over Brownian trajectories leads to the standard Schr\"{o}dinger equation,
while path integrals over \textquotedblleft\ skipping\textquotedblright\ L\'{e%
}vy trajectories lead to the FSE \cite{Laskin2}. Experimental schemes have
been proposed for realization of FSE in condensed-matter settings \cite%
{FSE11,excitons} and in optical media \cite{FSE12}, where the effective
fractional diffraction may be realized in cavities including optical filters
and phase masks \cite{early1,early2}. Actually, optical cavities offer a
testbed to explore the propagation dynamics and eigenmodes of optical beams
governed by linear and nonlinear FSEs. Following this approach, many
noteworthy properties have been revealed in the framework of FSE \cite%
{FSE13,FSE14,FSE15,FSE16,FSE17,FSE18,FSE19,FSE20,FSE21,FSE22,FSE23,FSE24,FSE25}%
. Adding nonlinear terms to FSE \cite{NLFSE1,NLFSE2}, recent publications
have predicted a variety of fractional solitons \cite%
{NLFSE3,NLFSE4,NLFSE5,NLFSE6,NLFSE7,NLFSE8,NLFSE9,NLFSE10,NLFSE11,NLFSE12,NLFSE13, NLFSE14,NLFSE15,NLFSE16,NLFSE17,NLFSE18,NLFSE19,NLFSE20,NLFSE21,LPF-Mal-Mih1,CQDai,Zeng, Yingji_latest,LPF-Mal-Mih2}%
.

In this work, we aim to construct circular soliton clusters in the model
combining fractional diffraction and competing CQ nonlinear terms. The paper
is organized as follows. The model, including a discussion of its
implementation in optics, and numerically found fundamental solitons, is
introduced in Sec. \ref{Sec II}, which is followed by identification of an
effective potential energy of the interaction between adjacent solitons, and
construction of necklace-shaped soliton clusters, in a semi-analytical form,
in Sec. \ref{Sec III}. Robust propagation of metastable soliton clusters
under the action of perturbations is additionally reported, by means of
systematic numerical simulations, in Sec. \ref{Sec IV}. The paper is
concluded by Sec. \ref{Sec V}.

\section{The model and fundamental solitons}

\label{Sec II}

We consider beam propagation along the $z$-axis in a nonlinear isotropic
medium with the CQ nonlinear correction to the refractive index, $n_{\mathrm{%
nonlin}}(I)=n_{2}I-n_{4}I^{2}$, where $I$ is the light intensity, $n_{2}$
and $n_{4}$ being coefficients accounting, respectively, for the cubic
self-focusing and quintic defocusing. The respective fractional NLSE, with
propagation distance $z$ and transverse coordinates $\left( x,y\right) $, is%
\begin{equation}
2ik_{0}\frac{\partial A}{\partial z}-\left( -\nabla _{\perp }^{2}\right)
^{\alpha /2}A+\frac{2k_{0}^{2}}{n_{0}}n_{\mathrm{nonlin}}A=0,  \label{NLFSE1}
\end{equation}%
where $A(z,x,y)$ is the local amplitude of the optical field, the intensity
being $I\equiv |A|^{2}$, $k_{0}=2\pi n_{0}/\lambda $ is\ the wavenumber,
which is determined by carrier wavelength $\lambda $, and $n_{0}$ is the
background refractive index. The fractional-diffraction operator in Eq. (\ref%
{NLFSE1}), with the corresponding L\'{e}vy index $\alpha $, belonging to
interval $1\leq \alpha \leq 2$, is defined as%
\begin{equation}
\left( -\nabla _{\perp }^{2}\right) ^{\alpha /2}A\left( x,y\right) =\mathcal{%
F}^{-1}\left[ \left( k_{x}^{2}+k_{y}^{2}\right) ^{\alpha /2}\mathcal{F}A(x,y)%
\right] ,  \label{operator}
\end{equation}%
\cite{JMP51-062102,JMP54-012111,CMA71}, where $\mathcal{F}$ and $\mathcal{F}%
^{-1}$ are operators of the direct and inverse Fourier transform, $k_{x,y}$
being wavenumbers conjugate to transverse coordinates $\left( x,y\right) $.
By means of rescaling, $\Psi (\zeta ,\xi ,\eta )=\sqrt{n_{4}/n_{2}}A(z,x,y)$%
, $\zeta =\left( k_{0}n_{2}^{2}/n_{0}n_{4}\right) z$, and $\left( \xi ,\eta
\right) =\left( 2k_{0}^{2}n_{2}^{2}/n_{0}n_{4}\right) ^{1/\alpha }\left(
x,y\right) $, Eq. (\ref{NLFSE1}) is cast in the normalized form,

\begin{equation}
i\frac{\partial \Psi }{\partial \zeta }-\left( -\nabla _{\perp }^{2}\right)
^{\alpha /2}\Psi +\left\vert \Psi \right\vert ^{2}\Psi -\left\vert \Psi
\right\vert ^{4}\Psi =0.  \label{NLFSE2}
\end{equation}%
Equation (\ref{NLFSE1}) with $\alpha \leq 1$ and cubic-only self-focusing
gives rise to the wave collapse \cite{NLFSE21}, which destabilizes solitons;
however, the quintic self-defocusing term suppresses the instability and
makes it possible to construct stable soliton states for $\alpha \leq 1$ as
well \cite{LPF-Mal-Mih1}.

In the linear limit, Eq. (\ref{NLFSE2}) models the free transmission of
optical beams under the action of fractional diffraction. Experimental
setups which realize this propagation regime were proposed in Refs. \cite%
{FSE12} and \cite{FSE14}. In particular, the former setting is designed as a
Fabry-Perot resonator, with two convex lenses and two phase masks inserted
into it. The first lens converts the input beam into the Fourier (\textit{%
dual}) space, then a central phase mask, whose position defines the position
of the system's \textit{Fourier plane}, performs the transformation of the
beam in the dual space, as per Eq. (\ref{operator}), and, eventually, the
second lens converts the output beam back from the dual domain into the real
(coordinate) one. Thus, the fractional diffraction is effectively executed
in the Fourier representation of the field amplitude, and the linear version
of Eq. (\ref{NLFSE2}) governs the circulation of the beam in the cavity,
averaged over many cycles. As a result, the linearized version of Eq. (\ref%
{NLFSE2}) gives rise to conical diffraction of 1D and 2D input Gaussian
beams \cite{FSE14,FSE19}.

The usual cubic nonlinearity can be incorporated in the setup by inserting a
piece of a Kerr material in the cavity. To maintain the effectively local
form of the nonlinearity, the material should be inserted between an edge
mirror and the lens closest to it, where the beam's propagation takes place
in the spatial domain (rather than in the dual one). The CQ terms may be
similarly induced by a sample of an optical material combining the
self-focusing and defocusing cubic and quintic nonlinearities, whose
absorption is low enough to be neglected. This may be, as mentioned above,
carbon disulfide in the liquid form, which was used for the creation of
stable (2+1)-dimensional spatial solitons \cite{PRL110-013901}, quasi-stable
(2+1)-dimensional optical solitons with embedded vorticity \cite%
{PRA93-013840}, and, very recently, of quasi-stable necklace arrays of
optical beams with imprinted vorticity \cite{Albert}. Another appropriate CQ
material is a colloidal suspension of metallic nanoparticles, in which
strengths of cubic and quintic nonlinearities may be accurately adjusted by
selecting the size and concentration of the particles \cite{Cid}.

Proceeding to the consideration of the full nonlinear equation (\ref{NLFSE2}%
), we first address stationary solutions with a real propagation constant, $%
\beta $:
\begin{equation}
\Psi \left( \zeta ,\xi ,\eta \right) =\psi \left( \xi ,\eta \right)
e^{i\beta \zeta },  \label{Solu}
\end{equation}%
where complex function $\psi (\xi ,\eta )$ obeys the equation%
\begin{equation}
-\left( -\nabla _{\perp }^{2}\right) ^{\alpha /2}\psi +\left\vert \psi
\right\vert ^{2}\psi -\left\vert \psi \right\vert ^{4}\psi -\beta \psi =0.
\label{NLFSE3}
\end{equation}%
\ Further, the complex function is substituted in the Madelung's form, $\psi
(\xi ,\eta )=U(\xi ,\eta )e^{i\phi (\xi ,\eta )}$, with real amplitude $%
U(\xi ,\eta )$ and phase $\phi (\xi ,\eta )$. In polar coordinates $%
(r,\theta )$, related to the underlying Cartesian coordinates as usual, $\xi
=r\cos \theta $, $\eta =r\sin \theta $, soliton solutions to Eq. (\ref%
{NLFSE2}) with integer embedded vorticity $s$ are looked for as \cite%
{LPF-Mal-Mih2}
\begin{equation}
\psi (r,\theta )=U(r)e^{is\theta }.  \label{psiU}
\end{equation}%
The self-trapped fundamental states, with $s=0$, are known as \textit{Townes
solitons} of the conventional 2D NLSE with the cubic-only term. In 1D, exact
fundamental solitons of the equation with the quintic-only self-focusing
term also play the role of solitons of the Townes type \cite{Salerno,Shamriz}%
. These solitons are unstable because the same equations admit the critical
collapse \cite{Fibich}. In the case of the CQ nonlinearity, the fundamental
solitons, in 2D and 1D settings alike, realize the system's ground state
(which is missing in the presence of the collapse).

The self-trapping of the solitons implies that $U(r)$ exponentially decays
at $r\rightarrow \infty $. Using the same argument as recently applied to
the FSE with operator $\left( -\partial ^{2}/\partial x^{2}\right) ^{\alpha
/2}$ \cite{NLFSE21}, \textit{viz}., the consideration of the variation of
the integral expression, which defines the nonlocal operator (\ref{operator}%
), with respect to rescaling of coordinates $x$ and $y$, it is
straightforward to conclude that a general form of the exponentially
decaying tails of the soliton is%
\begin{equation}
U(r)\propto \exp \left[ -\left( \frac{2}{C_{\alpha }}\beta \right)
^{1/\alpha }r\right] ,  \label{exp}
\end{equation}%
where the positive constant $C_{\alpha }$ is not universal, except for $%
C_{\alpha =2}=1$. Note also that the proportionality multiplier, dropped in
Eq. (\ref{exp}), includes a power-law pre-exponential factor $\sim
r^{-\gamma _{\alpha }}$ with some constant $\gamma _{\alpha }>0$ (obviously,
$\gamma _{\alpha =2}=1/2$).

The power (alias integral norm) and $z$-component of the angular momentum of
the vortex are%
\begin{equation}
P=\iint \left\vert \psi \left( \xi ,\eta \right) \right\vert ^{2}d\xi d\eta ,
\label{Energy}
\end{equation}%
\begin{equation}
L_{z}=\frac{\mathbf{e}_{z}}{2i}\cdot \iint \mathbf{\rho }\times \left( \psi
^{\ast }\nabla \psi -\psi \nabla \psi ^{\ast }\right) d\xi d\eta ,
\label{AM}
\end{equation}%
where $\mathbf{\rho }=\left( \xi ,\eta \right) $ is the vector of the
integration coordinates, and $\mathbf{e}_{z}$ is a unit vector in the
propagation direction. The Hamiltonian of Eq. (\ref{NLFSE2}) is
\begin{equation}
H=\iint \left[ \Psi ^{\ast }\left( -\nabla _{\perp }^{2}\right) ^{\alpha
/2}\Psi -\frac{1}{2}\left\vert \Psi \right\vert ^{4}+\frac{1}{3}\left\vert
\Psi \right\vert ^{6}\right] d\xi d\eta .  \label{Ham}
\end{equation}%
Note that, with regard to the definition given by Eq. (\ref{operator}),
expression (\ref{Ham}) produces a real Hamiltonian \cite{Laskin1}.

\begin{figure}[tbph]
\centering\includegraphics[width=8cm]{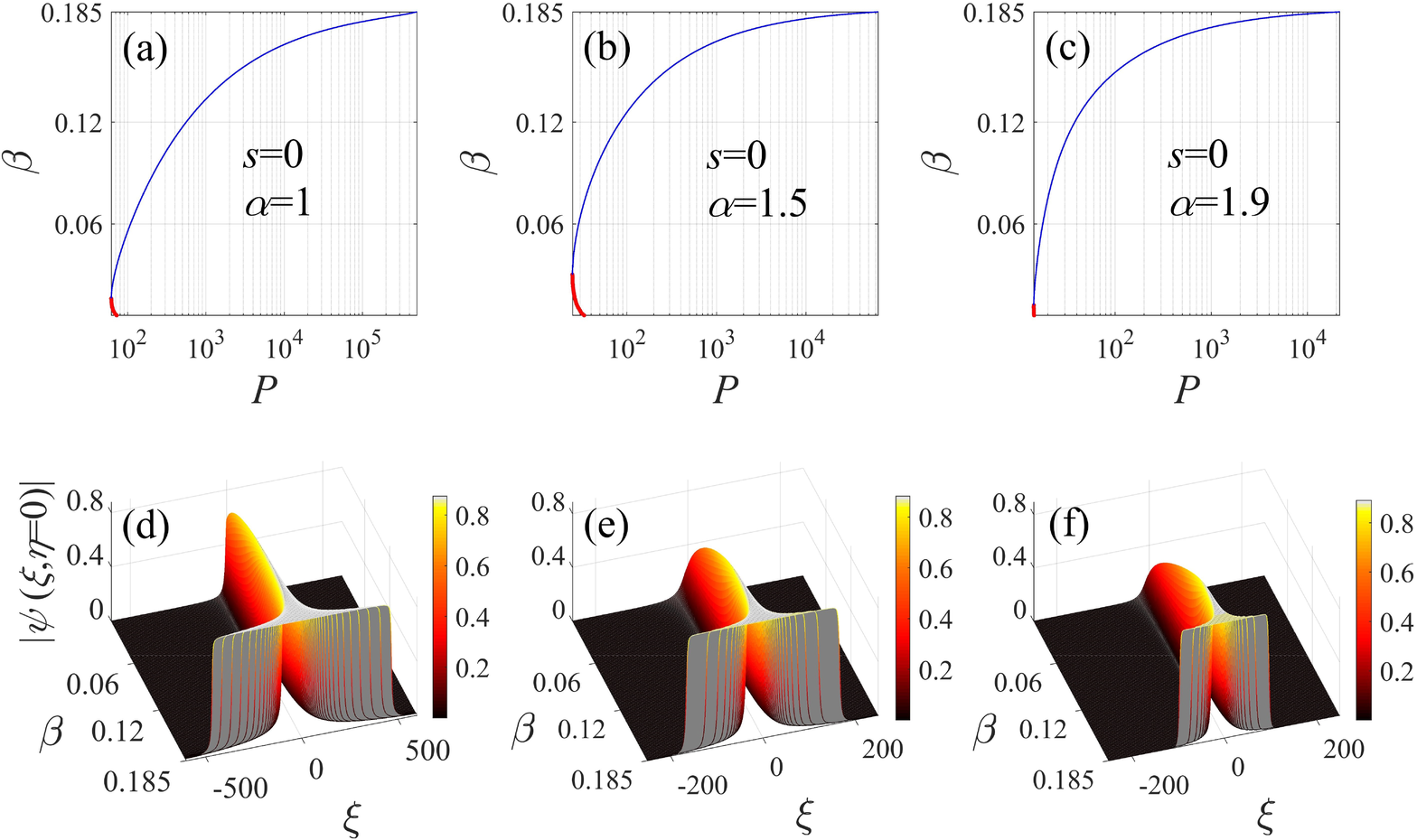}
\caption{The propagation constant of the fundamental solitons (with $s=0$)
versus their integral power (shown on the logarithmic scale), for different
values of the L\'{e}vy index: (a) $\protect\alpha =1$; (b) $\protect\alpha %
=1.5$; (c) $\protect\alpha =1.9$. Thin blue and thick red curves denote
stable and unstable solutions, respectively. Profiles of the cross-section, $%
\left\vert \protect\psi \left( \protect\xi ,\protect\eta =0\right)
\right\vert $, of the fundamental solitons for the propagation constant
taking values $0.006\leq \protect\beta \leq 0.185$, with step $\Delta
\protect\beta =0.001$, are depicted in panels (d) for $\protect\alpha =1$,
(e) for $\protect\alpha =1.5$, and (f) for $\protect\alpha =1.9$.}
\label{figure1}
\end{figure}

Families of fundamental solitons with different values of the L\'{e}vy index
$\alpha $ were numerically generated by dint of the
Newton-conjugate-gradient method \cite{NCG,Book1}. We display dependences of
their propagation constant $\beta $ on power $P$ for different values of the
L\'{e}vy index in Fig. \ref{figure1}. It is seen that two branches of the $%
\beta (P)$ curves merge at points with the vertical tangent, $d\beta /dP=0$.
Top ($d\beta /dP>0$) and bottom ($d\beta /dP<0$) branches are, severally,
stable and unstable, as established below by the linear-stability analysis
and direct simulations. These conclusions are correctly predicted by the
celebrated Vakhitov-Kolokolov criterion \cite{VK,Fibich}. Note that this
criterion is irrelevant for the stability of vortex solitons with $s\geq 1$
against azimuthal instabilities that may split the ring-shaped vortex modes
\cite{LPF-Mal-Mih2}. The point with $d\beta /dP=0$ corresponds to a
threshold (minimum) value of the total power, $P_{\mathrm{th}}^{(s=0)}$,
necessary for the existence of the fundamental solitons. In the cases shown
in Fig. \ref{figure1} the numerically found threshold values for the
fundamental solitons are $P_{\mathrm{th}}^{(s=0)}(\beta =0.016)|_{\alpha
=1}\approx 61.764$, $P_{\mathrm{th}}^{(s=0)}(\beta =0.03)|_{\alpha
=1.5}\approx 24.734$, and $P_{\mathrm{th}}^{(s=0)}(\beta =0.011)|_{\alpha
=1.9}\approx 14.271$. Further, the cross-section profiles of the fundamental
solitons, in the form of $\left\vert \psi \left( \xi ,\eta =0\right)
\right\vert $, are displayed in Figs. \ref{figure1}(d-f), for propagation
constants $\beta $ ranging from $0.006$ to $0.185$. As is common for models
with competing nonlinearities, the solitons develop a flat-top shape in the
limit of large powers.

The goal of this paper is to construct soliton clusters composed of
fundamental solitons in free space (without an external potential). Before
constructing the clusters, the stability of the fundamental solitons needs
to be established, which can be done in the framework of the linearization
of Eq. (\ref{NLFSE2}) for small perturbations. To this end, we searched for
the perturbed solutions of Eq. (\ref{NLFSE2}) in a form%
\begin{equation}
\Psi \left( \xi ,\eta ,\zeta \right) =e^{i\beta \zeta }\left[ \psi \left(
\xi ,\eta \right) +\epsilon u\left( \xi ,\eta \right) e^{\delta \zeta
}+\epsilon v^{\ast }\left( \xi ,\eta \right) e^{\delta ^{\ast }\zeta }\right]
,  \label{Perturbation}
\end{equation}%
where $\psi $ is the stationary\ solution with propagation constant $\beta $
defined as per Eq. (\ref{Solu}), the asterisk stands for the complex
conjugation, while $u\left( \xi ,\eta \right) $ and $v\left( \xi ,\eta
\right) $ are complex eigenmodes of perturbations, with an infinitesimal
amplitude $\epsilon $ \cite{Birnbaum}. Further, $\delta \equiv \delta
_{R}+i\delta _{I}$ is a complex eigenvalue of Eq. (\ref{LinearEigEqs}), $%
\delta _{R}$ being the linear-instability growth rate. Obviously, the
stability condition amounts to $\delta _{R}=0$, for all eigenvalues.

\begin{figure}[tbph]
\centering\includegraphics[width=8cm]{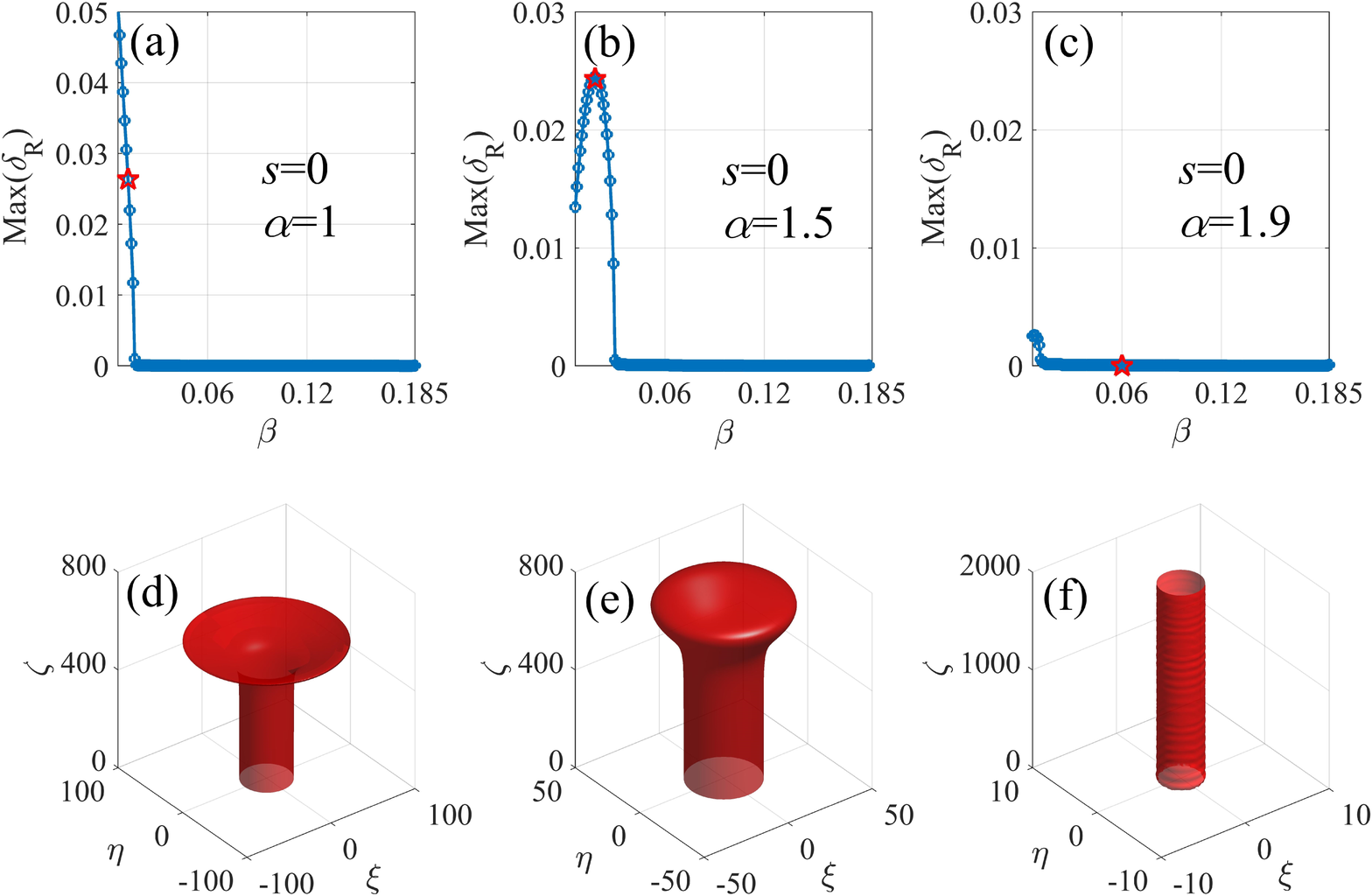}
\caption{Panels (a), (b), and (c) show the largest linear instability growth
rate of the fundamental-soliton families from Figs. \protect\ref{figure1}%
(a-c). Results of direct simulations of the evolution of particular
fundamental solitons, marked by red stars in the top row, are displayed by
isosurfaces drawn at a constant-intensity level, $\left\vert \Psi
\right\vert ^{2}=0.05$ in panels (d,e), and $\left\vert \Psi \right\vert
^{2}=0.50$ in (f). Panel (d): an unstable fundamental soliton, with $\protect%
\alpha =1$ and $\protect\beta =0.012$; (e) an unstable one, with $\protect%
\alpha =1.5$ and $\protect\beta =0.018$; (f) a stable fundamental soliton
with $\protect\alpha =1.9$ and $\protect\beta =0.06$.}
\label{figure2}
\end{figure}

The substitution of expression (\ref{Perturbation}) in Eq. (\ref{NLFSE2})
and subsequent linearization leads to the following eigenvalue problem:%
\begin{equation}
i\left(
\begin{array}{cc}
\mathit{L}_{11} & \mathit{L}_{12} \\
-\mathit{L}_{12}^{\ast } & -\mathit{L}_{11}^{\ast }%
\end{array}%
\right) \left(
\begin{array}{c}
u \\
v%
\end{array}%
\right) =\delta \left(
\begin{array}{c}
u \\
v%
\end{array}%
\right) ,  \label{LinearEigEqs}
\end{equation}%
with matrix elements $\mathit{L}_{11}=-\left( -\nabla _{\perp }^{2}\right)
^{\alpha /2}+2\left\vert \psi \right\vert ^{2}-3\left\vert \psi \right\vert
^{4}-\beta $ and $\mathit{L}_{12}=\psi ^{2}+2\left\vert \psi \right\vert
^{2}\psi ^{2}$. In the context of BEC, linearized equations for small
perturbations are usually called Bogoliubov-de Gennes equations \cite{Pit}.
Note that the choice of the ansatz for the perturbed solution in the form of
Eq. (\ref{Perturbation}), with the combination of amplitudes $u$ and $%
v^{\ast }$, is convenient because it leads to Eq. (\ref{LinearEigEqs}) for
two-component eigenfunctions which does not include complex conjugation.

The linear problem based on Eq. (\ref{LinearEigEqs}) was solved by means of
the Newton-conjugate-gradient method \cite{NCG,Book1}. Differently from the
conventional NLSE with the CQ nonlinearity, corresponding to $\alpha =2$ in
Eq. (\ref{NLFSE1}), in which the fundamental solitons are completely stable
\cite{Pego}, the fundamental solitons are found to be unstable in intervals $%
0.006<\beta \leq 0.016$ (for $\alpha =1$), see Fig. \ref{figure2}(a), $%
0.006<\beta \leq 0.030$ (for $\alpha =1.5$) in Fig. \ref{figure2}(b), and $%
0.006<\beta \leq 0.011$ (for $\alpha =1.9$) in Fig. \ref{figure2}(c). These
results corroborate that the fundamental-soliton branches with a negative
slope, $d\beta /dP<0$, are linearly unstable, while the positive-slope ones
are stable, i.e., the stability of the fundamental solitons fully complies
with the Vakhitov-Kolokolov criterion. In Figs. \ref{figure2}(d) and \ref%
{figure2}(e), the results of direct simulations demonstrate that unstable
fundamental solitons at first keep their integrity, and then start to spread
out, see also movies Visualization 1 and Visualization 2 in Supplemental
Material. In Fig. \ref{figure2}(f), a linearly-stable fundamental soliton
robustly propagates, under the action of random perturbations with a $5\%$
relative amplitude, over the distance of $\zeta =2000$, which is tantamount
to $\sim 50$ diffraction lengths of this soliton.

\section{The interaction energy and construction of soliton clusters}

\label{Sec III}

The necklace-shaped soliton cluster of radius $R$ is constructed as a
superposition of $N$ identical fundamental solitons with a superimposed
phase distribution carrying the integer global vorticity $M$. Thus, the
initial field distribution is set as%
\begin{equation}
\Psi \left( \zeta =0\right) =\exp \left( iM\theta \right)
\sum\limits_{n=1}^{N}U_{0}(\left\vert \mathbf{r}-\mathbf{r}_{n}\right\vert ),
\label{SC}
\end{equation}%
where\ $U_{0}$\ is a stable fundamental soliton, and $\mathbf{r}_{n}=\{R\cos
(2\pi n/N),R\sin (2\pi n/N)\}$ is the position of the center of the $n$-th
soliton in the cluster. The evolution of the pattern is governed by
interaction forces between adjacent solitons, which, in turn, depend on the
phase shift and separation between them. Note that ansatz (\ref{SC})
includes a linear phase distribution with respect to the angular coordinate,
$\theta $, rather than a staircase phase structure, because the former one
is more likely to form a quasi-stable structure \cite{Crasovan-PRE67-046610}.

As concerns the input necessary for the creation of soliton clusters
introduced in the form of Eq. (\ref{SC}), in the experiment it can be put
together as a set of several parallel beams placed along the ring, while a
vortex phase plate may be used to imprint the phase pattern, which
represents the overall vorticity [$M$ in Eq. (\ref{SC})], onto the
multi-beam cluster.

Here, we emphasize the difference of physical properties between the vortex
soliton and necklace-shaped soliton cluster. Vortex solitons in Ref. \cite%
{LPF-Mal-Mih2} are stationary solutions of Eq. (\ref{NLFSE2}), which may be
stable or unstable. However, necklace-shaped soliton clusters, discussed in
this paper, are not stationary solutions of Eq. (\ref{NLFSE2}). They are
composed of stable fundamental solitons of Eq. (\ref{NLFSE2}), placed along
a ring. While Eq. (\ref{NLFSE2}) produces no completely stable
necklace-shaped soliton clusters, they can be found in a quasi-stable form.
Their existence is predicted by minimization of the effective potential of
interaction between adjacent solitons in the cluster, cf. Ref. \cite%
{Kartashov-Malomed-Torner}.

Since the necklace may be considered as an azimuthally modulated vortex
soliton, one may expect that necklace clusters can be maintained by Eq. (\ref%
{NLFSE2}) if their integral power exceeds the known threshold value, $P_{%
\mathrm{th}}^{(s=1)}$ \cite{LPF-Mal-Mih2}, above which this equation gives
rise to stable vortex solitons. This circumstance is elaborated below, see
Eq. (\ref{N}).

\begin{figure}[tbph]
\centering\includegraphics[width=8cm]{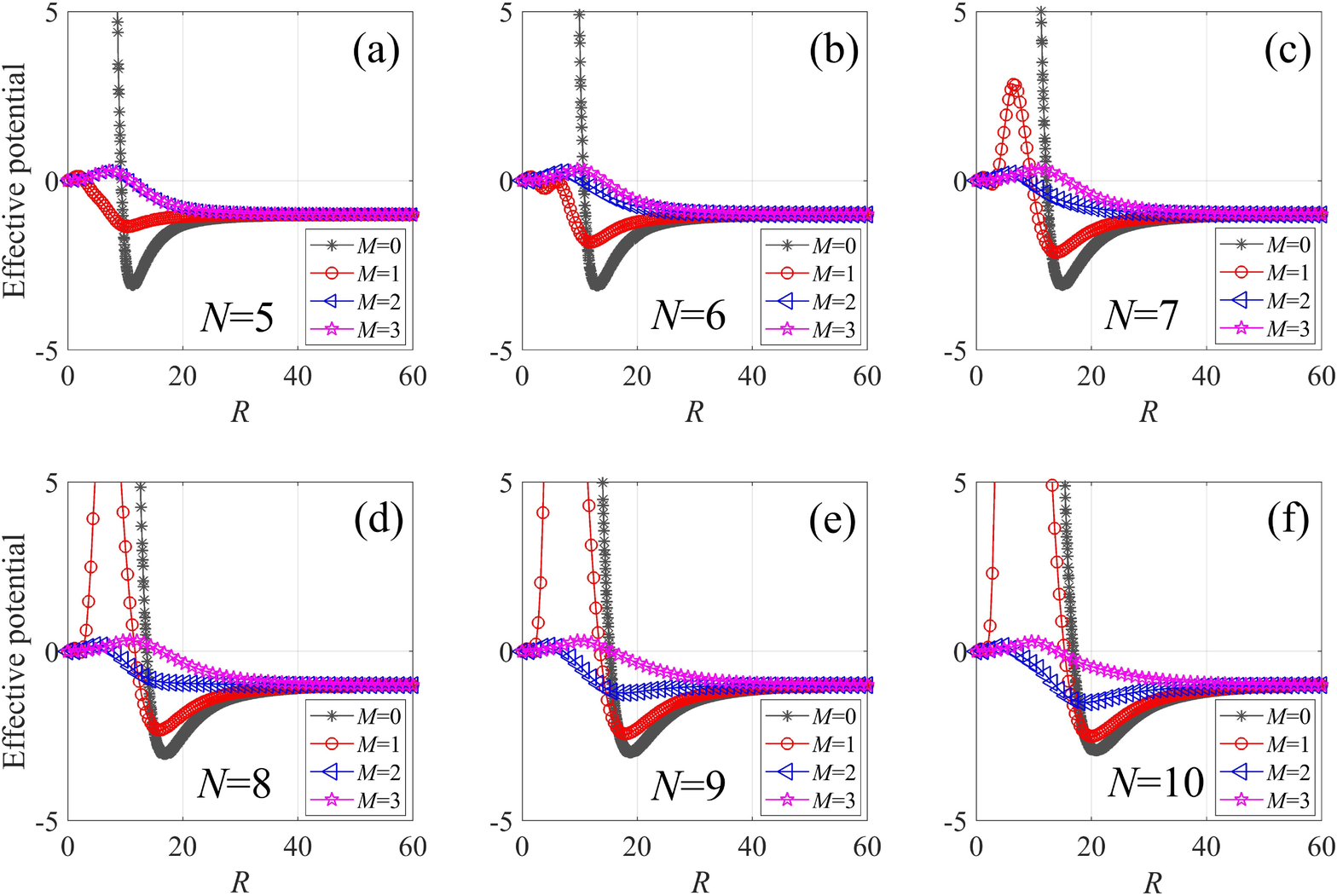}
\caption{The effective potential given by ansatz (\protect\ref{SC}),
numerically computed as per Eq. (\protect\ref{pot}), with different values
of vorticity $M$ and number of individual fundamental solitons, $N$, vs. the
cluster's radius, $R$, for (a) $N=5$, (b) $N=6$, (c) $N=7$, (d) $N=8$, (e) $%
N=9$, and (f) $N=10$. Here, the L\'{e}vy index is $\protect\alpha =1.5$, and
the integral power of each individual soliton is fixed to be $P_{\mathrm{sol}%
}=84.1$.}
\label{figure3}
\end{figure}

\begin{table}[tbph]
\centering%
\begin{tabular}{|c|c|c|c|c|c|c|}
\hline
& $N=5$ & $N=6$ & $N=7$ & $N=8$ & $N=9$ & $N=10$ \\ \hline
$M=0$ & $11.25$ & $13.05$ & $14.90$ & $16.85$ & $18.80$ & $20.75 $ \\ \hline
$M=1$ & $10.35$ & $11.90$ & $13.85$ & $15.90$ & $17.95$ & $20.00$ \\ \hline
\end{tabular}
\caption{The values of $R_{\min }$, at which the effective potential (%
\protect\ref{pot}) of the necklace-cluster ansatz (\protect\ref{SC}), with
vorticities $M=0$ and $1$, attains a minimum, for different numbers $N$ of
individual fundamental solitons in the ansatz, see Fig. \protect\ref{figure3}%
.}
\label{table1}
\end{table}

The force of the interaction between fundamental solitons with phase
difference $\chi $, separated by large distance $L$, is determined by an
effective interaction potential, which is determined by the integral of
overlapping between an exponentially decaying tail of each soliton and the
core of its mate. Taking into regard the general asymptotic form of the
tails, given by Eq. (\ref{exp}), we conclude that the potential of the
soliton-soliton interaction may be estimated, as per Ref. \cite{1998}, as%
\begin{equation}
W(L)\simeq -W_{0}\cos \chi \cdot \exp \left[ -\left( \frac{2}{C_{\alpha }}%
\beta \right) ^{1/\alpha }L\right] ,  \label{W}
\end{equation}%
where $W_{0}$ is a positive factor, which may be a slowly decaying function
of $L$ (for $\alpha =2$, $W_{0}=\mathrm{const}\cdot L^{-1/2}$), while factor
$\cos \chi $ is a universal one \cite{1998}. For the ansatz defined by Eq. (%
\ref{SC}), one has $L=2R\sin \left( \pi /N\right) $, $\chi =2\pi M/N$.

For the whole necklace cluster, the scaled potential, normalized to the
energy of the same set of infinitely separated (non-interacting) solitons,
may be defined as
\begin{equation}
\mathrm{Effective~potential}=H(R)/H(R\rightarrow \infty ),  \label{pot}
\end{equation}%
where $H$ is the full Hamiltonian (\ref{Ham}). It can be computed
numerically as a function of $R$ for given vorticity $M$ and number of
individual \textquotedblleft\ bids in the necklace\textquotedblright\ , i.e.,
individual fundamental solitons, $N $, by substituting ansatz (\ref{SC}) in
integral expression (\ref{Ham}).

First, we address the case of $\alpha =1.5$. In this case, Eq. (\ref{NLFSE2}%
) maintains stable solitons with vorticity $s=1$ if the integral power
exceeds the above-mentioned threshold value, $P_{\mathrm{th}}^{(s=1)}\approx
383$, see Table 1 in Ref. \cite{LPF-Mal-Mih2}. Therefore, a set of
fundamental solitons, each taken (for instance) with $P_{\mathrm{sol}}=84.1$
and $\beta =0.12$, may have a chance to build robust clusters with $M=1$ and
\begin{equation}
N>P_{\mathrm{th}}^{(s=1)}/P_{\mathrm{sol}}\approx 4.6,  \label{N}
\end{equation}%
which actually implies $N\geq 5$.

In Fig. \ref{figure3}, the dependence of the effective potential (\ref{pot})
on the cluster's radius $R$ is presented for different integer values of $M$
and $N$. These results reveal that the necklace clusters with vorticities $%
M=0$ and $1$ exhibit visible local minima of the interaction potential at a
specific value of the cluster's radius, $R_{\min }$. These numerically
identified values for $M=0$ and $1$ are summarized in Table \ref{table1} for
different numbers of solitons in the cluster, $N$. The results demonstrate
that $R_{\min }$ increases with the increase of $N$. For the clusters with $%
M=2$, local energy minima are observed in Figs. \ref{figure3}(e,f) for $N=9$
and $10$, but, according to data from Table 1 in Ref. \cite{LPF-Mal-Mih2},
the respective values of the total power of initial cluster (\ref{SC}) are
lower than the corresponding stability-threshold value for the axisymmetric
vortex soliton, cf. Eq. (\ref{N}). Therefore, the minima of the potential
energy for $M=2$ cannot produce stable necklace patterns. Lastly, Fig. \ref%
{figure3} does not produce local energy minima for the clusters with $M=3$.

\begin{figure}[tbph]
\centering\includegraphics[width=8cm]{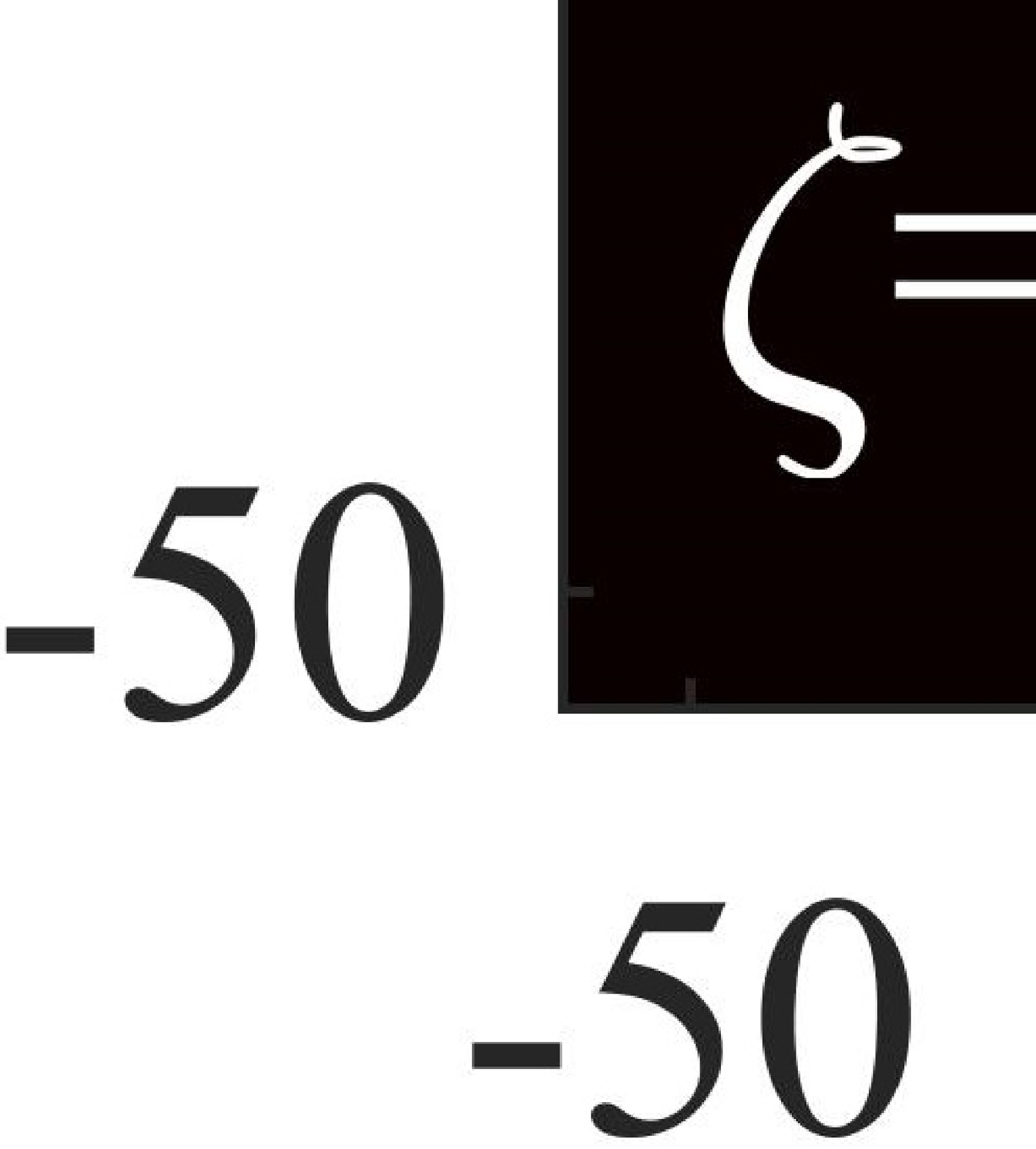}
\caption{Different scenarios of the evolution of initial necklace soliton
clusters (\protect\ref{SC}), built of $N=5$ fundamental solitons at $\protect%
\alpha =1.5$ (in the absence of random perturbations). (a) Gradual fusion
into a large fundamental soliton with initial mean angular velocity $\protect%
\omega _{0}=$ $0$ [given by Eq. (\protect\ref{MeanOmega}) at $\protect\zeta %
=0$], at $M=0$, $R=11.25$. (b) Expansion and rotation for $M=2$, $R=10.35$,
and $\protect\omega _{0}=$ $0.0128$. (c) Periodic oscillations and rotation
for $M=1$, $R=13$, and $\protect\omega _{0}=$ $0.005$. (d) Propagation of
the rotating cluster with a nearly constant radius, for $M=1$, $R=10.35$,
and $\protect\omega _{0}=$ $0.0071$. Local intensity $\left\vert \Psi
\right\vert ^{2}$ is shown at different values of propagation distances $%
\protect\zeta $, which correspond to red circles in the $R(\protect\zeta )$
dependences in the top row of Fig. \protect\ref{figure5}.}
\label{figure4}
\end{figure}

To check the predictions produced by the effective potential, we numerically
simulated the propagation of the necklace clusters built of $N=5$
fundamental solitons [see Fig. \ref{figure3}(a)] by\ means of the split-step
Fourier method. In the course of the simulations, the evolution of the
cluster's mean radius $R\left( \zeta \right) $, mean angular velocity $%
\omega \left( \zeta \right) $, and depth of the azimuthal modulation $\Theta
\left( \zeta \right) $ was monitored. These quantities were defined as
follows:%
\begin{equation}
R\left( \zeta \right) \equiv P^{-1}\iint \sqrt{\xi ^{2}+\eta ^{2}}\left\vert
\psi \left( \xi ,\eta \right) \right\vert ^{2}d\xi d\eta ,  \label{MeanR}
\end{equation}%
\begin{equation}
\omega \left( \zeta \right) \equiv L_{z}/I,I\equiv \iint \left( \xi
^{2}+\eta ^{2}\right) \left\vert \psi \left( \xi ,\eta \right) \right\vert
^{2}d\xi d\eta ,\omega _{0}=\omega \left( \zeta =0\right) ,
\label{MeanOmega}
\end{equation}%
and%
\begin{equation}
\Theta \left( \zeta \right) \equiv \frac{\left\vert \Psi \right\vert _{\max
}^{2}-\left\vert \Psi \right\vert _{\min }^{2}}{\left\vert \Psi \right\vert
_{\max }^{2}+\left\vert \Psi \right\vert _{\min }^{2}},  \label{DepthAM}
\end{equation}%
where $\left\vert \Psi \right\vert _{\max }^{2}$ and $\left\vert \Psi
\right\vert _{\min }^{2}$ are the maximum and minimum values of $\left\vert
\Psi \right\vert ^{2}$, as functions of $\theta $,\ along the circumference
at which the solution has a radial maximum in each cluster's configuration.

\begin{figure}[tbph]
\centering\includegraphics[width=9cm]{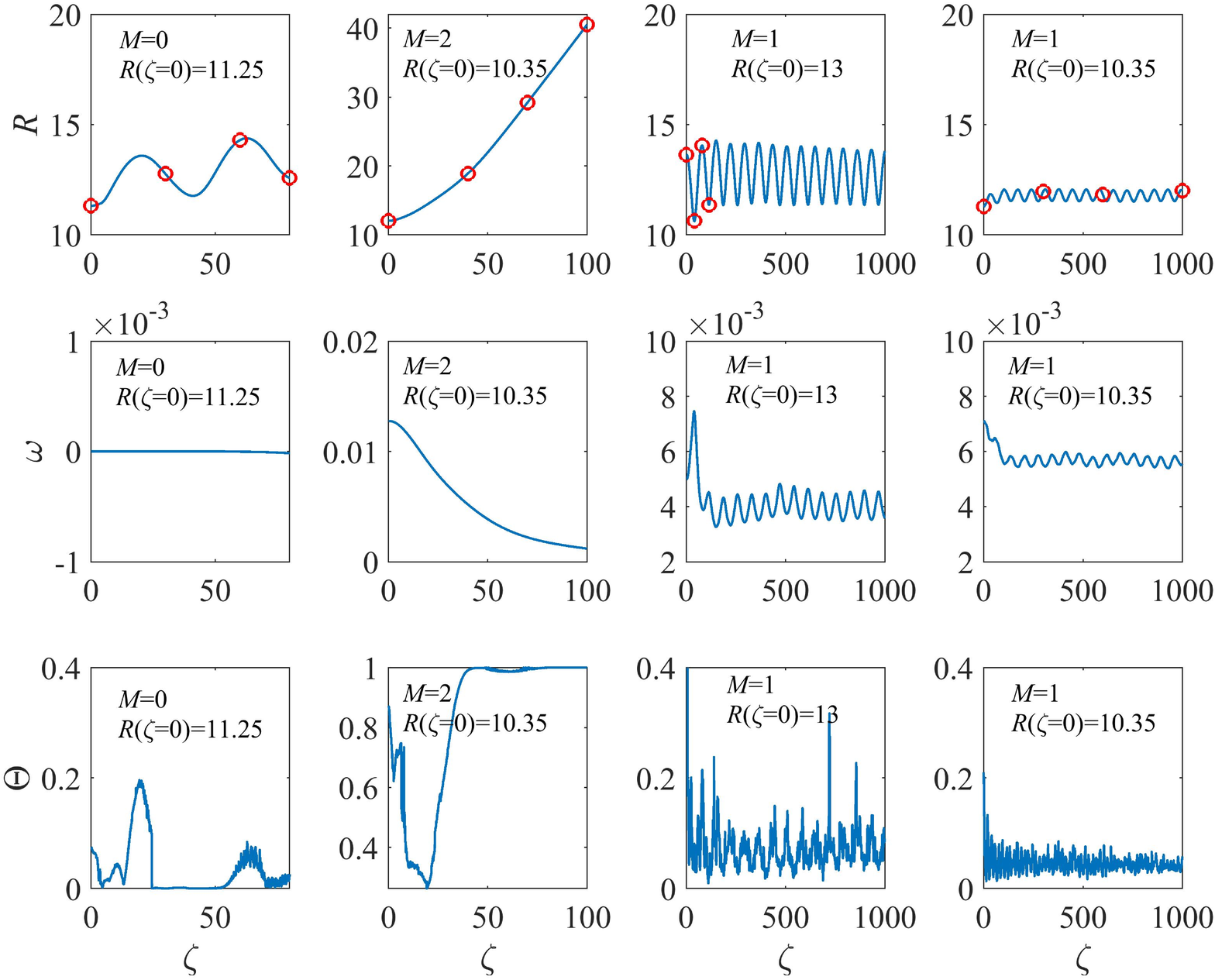}
\caption{The evolution of the cluster's mean radius (top row), mean angular
velocity (middle row), and depth of the azimuthal modulation (bottom row),
which are defined as per Eqs. (\protect\ref{MeanR}), (\protect\ref{MeanOmega}%
), and (\protect\ref{DepthAM}), respectively. The columns, from left to
write, correspond, respectively, to rows (a) -- (d) in Fig. \protect\ref%
{figure4}.}
\label{figure5}
\end{figure}

Results produced by the simulations are presented in Fig. \ref{figure4}. For
$M=0$, interactions between adjacent fundamental solitons are attractive,
according to Eq. (\ref{W}), therefore the multi-soliton cluster fuses into a
fundamental state, as shown in the top row of Fig. \ref{figure4}, even for
the initial radius taken as $R=R_{\min }$. On the other hand, for $M=2$, the
soliton-soliton interactions, with $\chi =\pi $, are repulsive, thus giving
rise to a resulting repulsive force acting on each individual soliton in the
radial direction. For this reason, the effective potential energy (\ref{pot}%
) has no local minimum in this case, see Fig. \ref{figure3}(a). As a result,
the cluster expands and rotates, as seen in Fig. \ref{figure4}(b). For $M=1$%
, i.e., $\chi =3\pi /5$, the interaction potential (\ref{W}) is weakly
attractive, but the net angular momentum of the circular cluster prevents
fusion into an axisymmetric vortex soliton. In this case, the cluster with
initial radius $R>R_{\min }$ performs quasiperiodic cycles of expansion and
shrinkage (in combination with rotation), as shown in Fig. \ref{figure4}(c).
The most essential finding is that the cluster with initial radius $R\approx
R_{\min }$ gives rise a quasi-stationary state, which performs persistent
rotation with minimal radial oscillations, see Fig. \ref{figure4}(d). This
mechanism of the effective stabilization of the rotating necklace cluster is
somewhat similar to that recently reported in the framework of the usual 2D
NLSE with the nonlinearity corresponding to quantum droplets \cite%
{Kartashov-Malomed-Torner}. In Fig. \ref{figure5}, the evolution of the
cluster's mean radius, cluster's mean angular velocity, and depth of the
azimuthal modulation is monitored, to display further details of the
variation of the cluster's structure and rotational speed in the course of
the propagations. The first panel in the middle row of Fig. \ref{figure5}
shows the case of the cluster's mean angular velocity equal to zero,
therefore it does not rotate. In the second panel in the middle row of Fig. %
\ref{figure5}, the cluster exhibits rotation with a decreasing angular
velocity. In the last two panels in the middle row of Fig. \ref{figure5},
the cluster's mean angular velocity fluctuates around non-decaying mean
values, which indicates that the respective clusters perform persistent
rotation with small radial oscillations in the transverse plane.

\section{Robust propagation of metastable soliton clusters under the action
of perturbations}

\label{Sec IV}

\begin{figure}[tbph]
\centering\includegraphics[width=9cm]{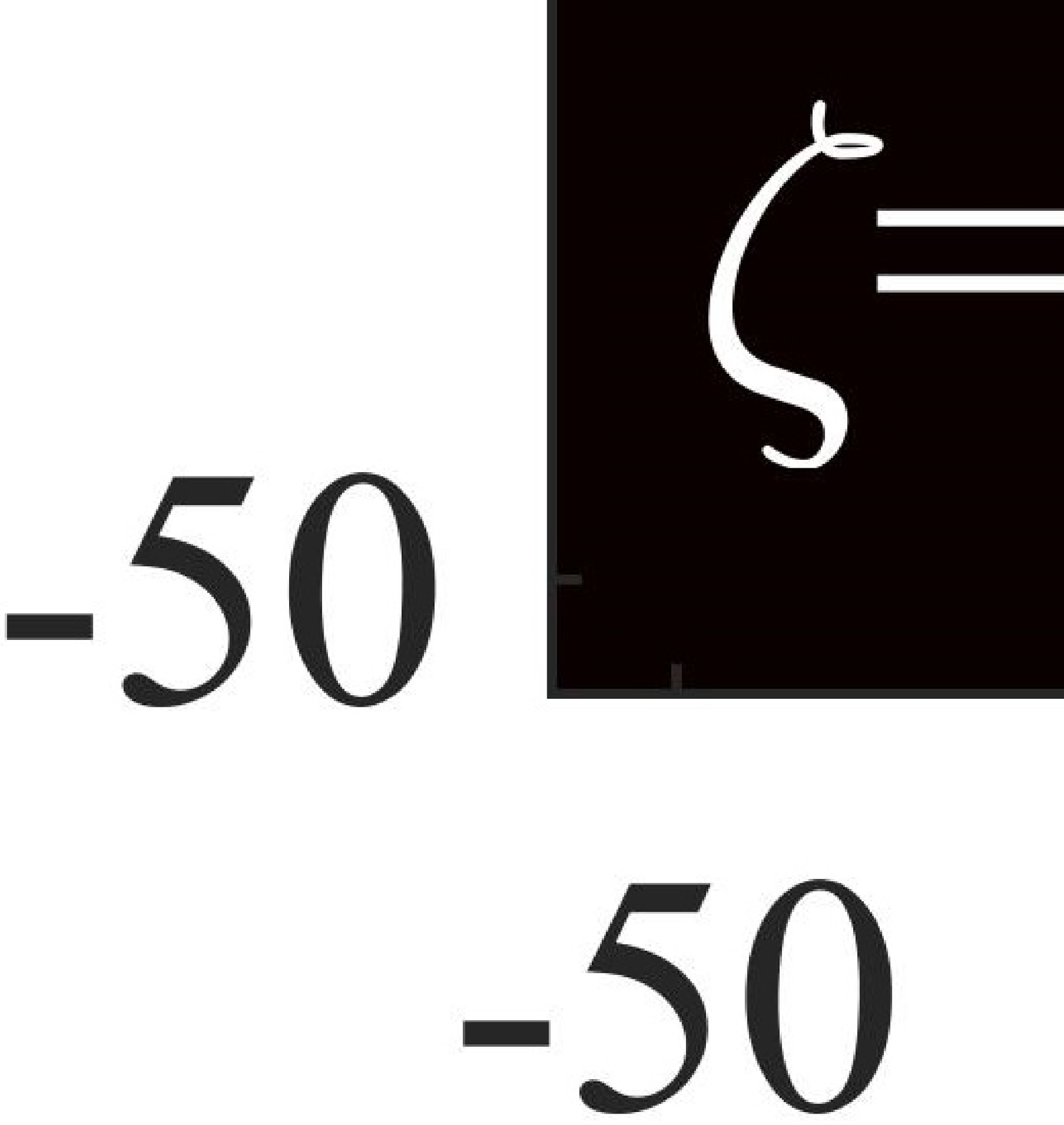}
\caption{Typical examples of the metastable propagation of the rotating
necklace-shaped soliton clusters built of $N$ fundamental solitons, with L%
\'{e}vy index $\protect\alpha =1.5$, initial radius $R=R_{\min }$, and
superimposed vorticity $M=1$, in the presence of random perturbations
produced by factor (\protect\ref{factor}): (a) $N=5$, $R=10.35$, and $%
\protect\omega _{0}=$ $0.0071$, (b) $N=6$, $R=11.9$, and $\protect\omega %
_{0}=$ $0.0058$, (c) $N=7$, $R=13.85$, and $\protect\omega _{0}=$ $0.0044$,
(d) $N=8$, $R=15.9$, and $\protect\omega _{0}=$ $0.0036$. The local
intensity $\left\vert \Psi \right\vert ^{2}$ is shown at different values of
propagation distances $\protect\zeta $, which correspond to red circles in
the $R(\protect\zeta )$ dependences in the top row of Fig. \protect\ref%
{figure7}.}
\label{figure6}
\end{figure}

\begin{figure}[tbph]
\centering\includegraphics[width=8cm]{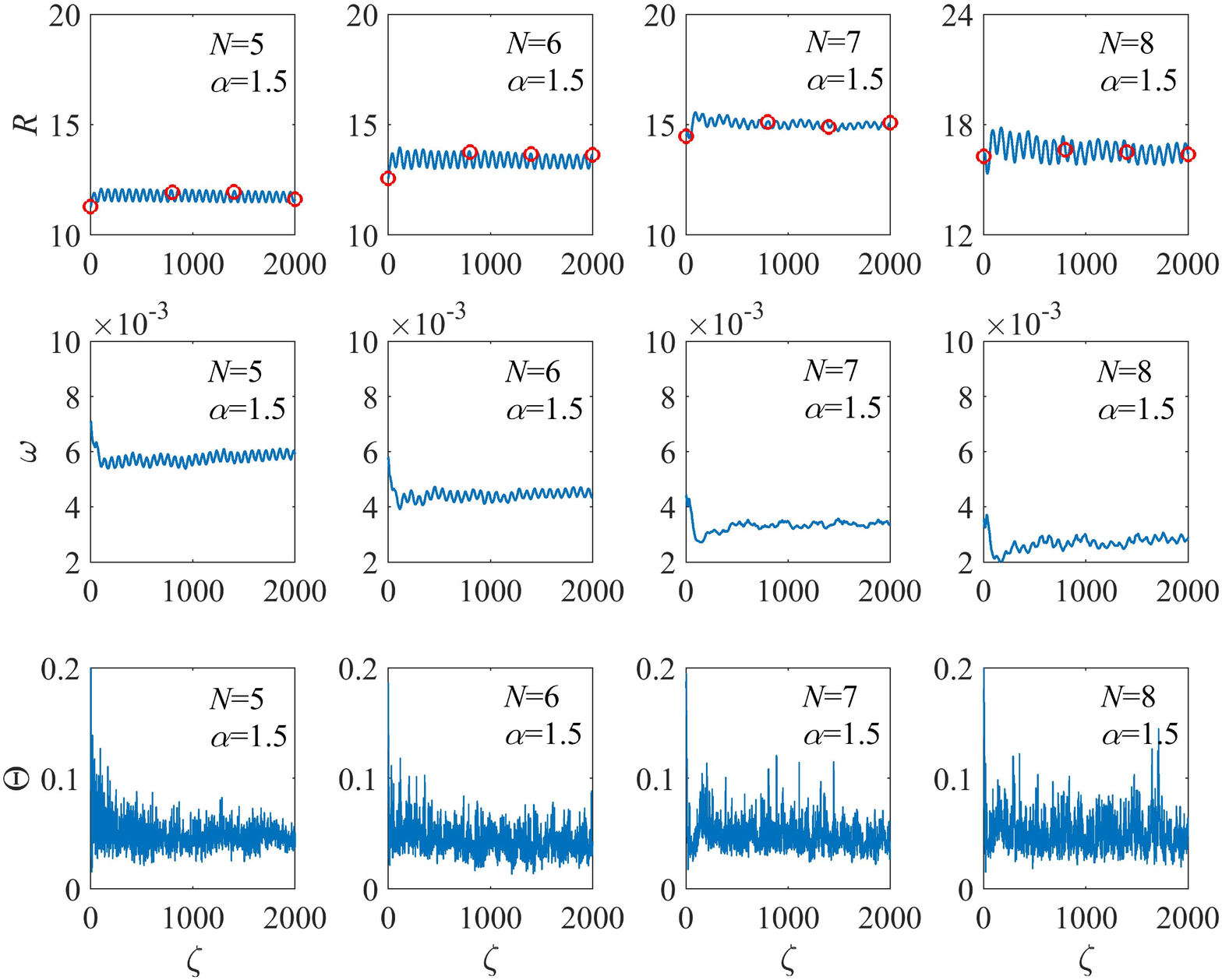}
\caption{The evolution of the cluster's mean radius (top row), mean angular
velocity (middle row), and depth of the azimuthal modulation (bottom row),
which are defined as per Eqs. (\protect\ref{MeanR}), (\protect\ref{MeanOmega}%
), and (\protect\ref{DepthAM}), respectively. The columns, from left to
write, correspond, respectively, to rows (a) -- (d) in Fig. \protect\ref%
{figure6}.}
\label{figure7}
\end{figure}

\begin{figure}[tbph]
\centering\includegraphics[width=8cm]{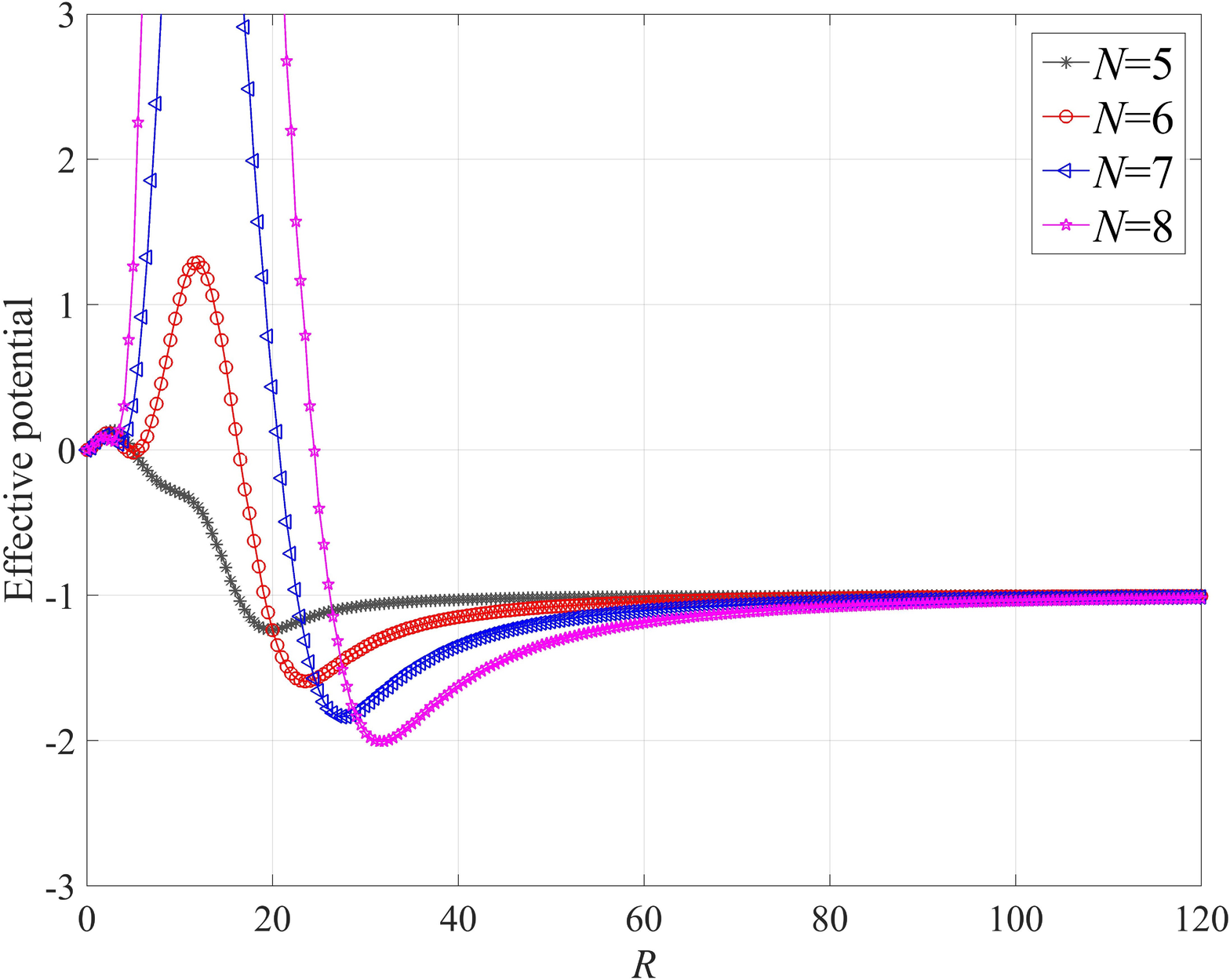}
\caption{The effective potential energy of the soliton clusters, composed of
$N$ fundamental solitons, with superimposed vorticity $M=1$, defined as per
Eq. (\protect\ref{pot}), vs. the cluster's radius $R$: (a) $N=5$, (b) $N=6$,
(c) $N=7$, (d) $N=8$. In this case, the L\'{e}vy index is $\protect\alpha =1$%
, and the integral power of all clusters is $P=332.35$.}
\label{figure8}
\end{figure}

\begin{figure}[tbph]
\centering\includegraphics[width=8cm]{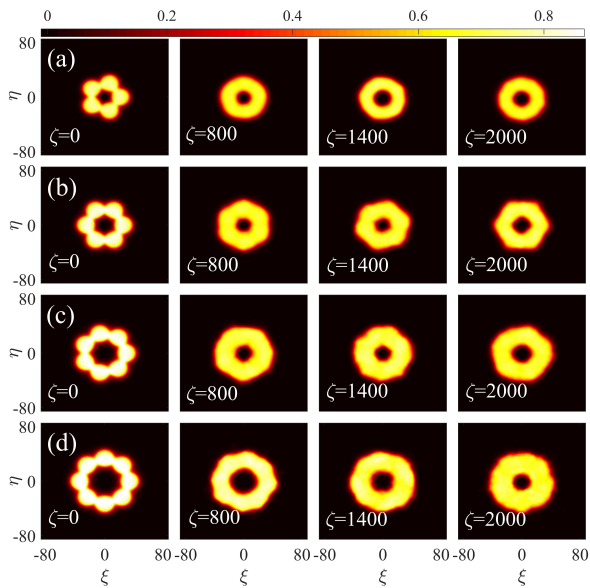}
\caption{Metastable propagation of the rotating soliton clusters with L\'{e}%
vy index $\protect\alpha =1$, initial radius $R=R_{\min }$, and vorticity $%
M=1$, in the presence of random perturbations imposed by factor (\protect\ref%
{factor}): (a) $N=5$, $R=19.8$, and $\protect\omega _{0}=$ $0.002$, (b) $N=6$%
, $R=23.55$, and $\protect\omega _{0}=$ $0.0015$, (c) $N=7$, $R=27.6$, and $%
\protect\omega _{0}=$ $0.0012$, (d) $N=8$, $R=31.6$, and $\protect\omega %
_{0}=$ $0.001$. The local intensity $\left\vert \Psi \right\vert ^{2}$ is
shown at different values of propagation distance $\protect\zeta $ that
correspond to red circles in the top row of Fig. \protect\ref{figure10}.}
\label{figure9}
\end{figure}

\begin{figure}[tbph]
\centering\includegraphics[width=9cm]{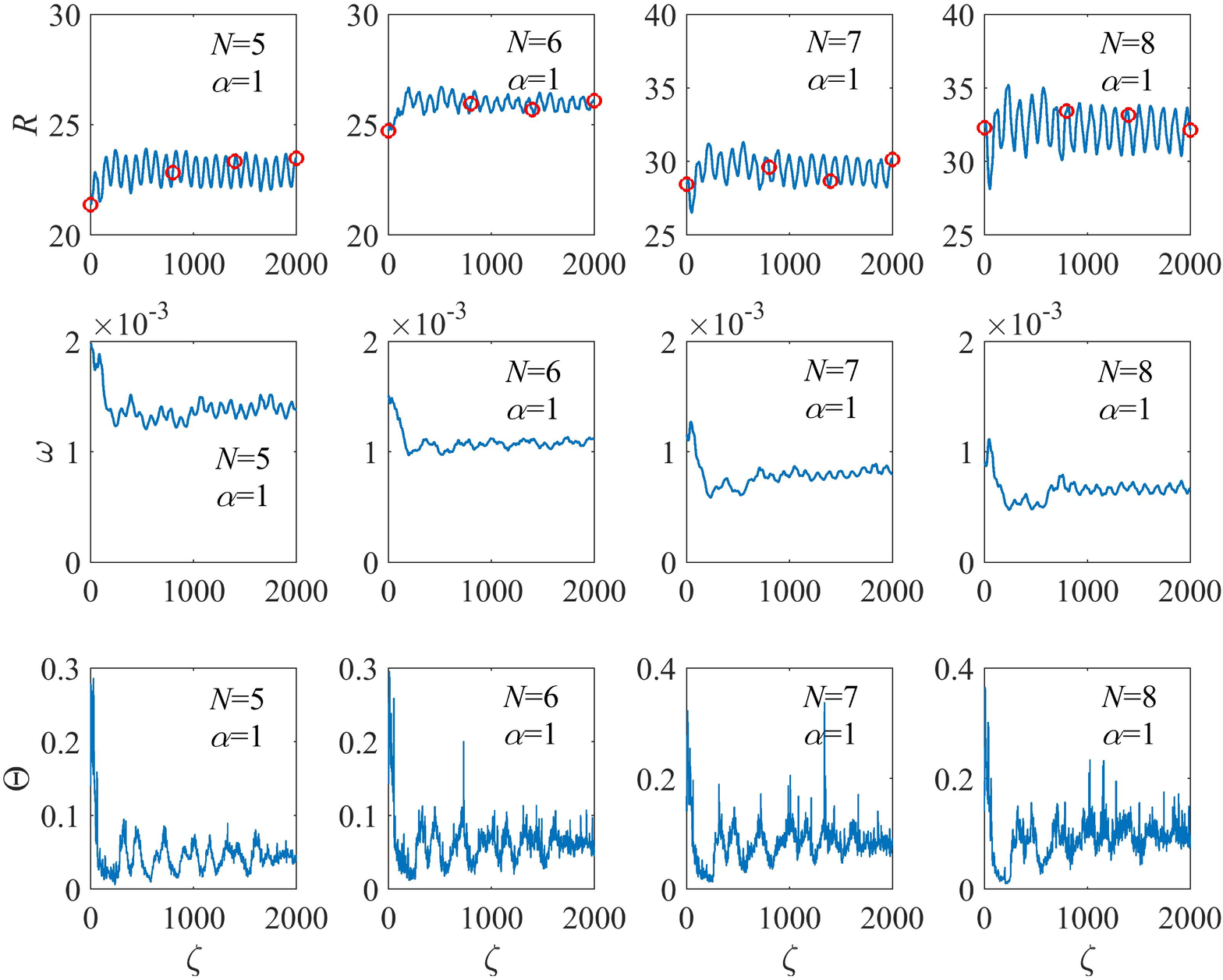}
\caption{The evolution of the cluster's mean radius (top row), mean angular
velocity (middle row), and depth of the azimuthal modulation (bottom row)
for different number of fundamental solitons $N$ in Fig. \protect\ref%
{figure9}. The columns, from left to write, correspond, respectively, to
rows (a) -- (d) in Fig. \protect\ref{figure9}.}
\label{figure10}
\end{figure}

To further elucidate the robustness of the soliton clusters with the
properly chosen initial radius, $R=R_{\min }$, we added random perturbation
to the initial profiles of the soliton cluster in Eq. (\ref{SC}),
multiplying it by
\begin{equation}
\mathrm{perturbing~factor}=[1+0.1(\rho _{1}+i\rho _{2})],  \label{factor}
\end{equation}%
where $\rho _{1}$ and $\rho _{2}$ are\ uniformly distributed random numbers
in the interval of $[-0.5,0.5]$. In Fig. \ref{figure6}, the simulations
demonstrate that the rotating necklace clusters, composed of different
numbers of fundamental solitons ($N=5,6,7$, and $8$), with initial radius $%
R=R_{\min }$, can survive, as metastable patterns, in the course of long
evolution, even under the action of relatively strong random perturbations.
In the Supplemental Material, we additionally display the evolution of the
clusters with vorticity $M=1$ and initial radius $R=R_{\min }$, for larger
numbers of the fundamental solitons, \textit{viz}., $N=9$ and $10$, see
movies Visualization 3 and Visualization 4, respectively. The results
indicate that the rotating necklace clusters remain robust against the
action of random perturbation, although exhibiting radial oscillations. In
the course of the propagations, the evolution of the cluster's mean radius
and mean angular velocity quasi-periodically oscillate with a small
amplitude, see the top and middle rows in Fig. \ref{figure7}. The depth of
the azimuthal modulation is also monitored, see the bottom row in Fig. \ref%
{figure7}, which indicates that the clusters do not fully fuse into the
usual solitons in the course of the long-distance propagation.

Next, we address the necklace-shaped clusters composed of the fundamental
solitons with L\'{e}vy index $\alpha =1$ [recall it is the critical value
for Eq. (\ref{NLFSE2}) with the cubic-only self-attraction, therefore it is
relevant to address this case too]. It was recently found that the
axisymmetric vortex soliton (not a cluster) with $s=1$ and $\alpha =1$ is
stable if its integral power exceeds the respective threshold value, $P>P_{%
\mathrm{th}}^{(s=1)}\approx 1553$ (see Table 1 in Ref. \cite{LPF-Mal-Mih2}).
Here, we construct the necklace patterns with vorticity $M=1$ as circular
sets of $N$ stable fundamental solitons, each one with power $P_{\mathrm{sol}%
}=332.35$ and propagation constant $\beta =0.105$. Thus, they have a chance
to be quasi-stable for $N>P_{\mathrm{th}}^{(s=1)}/P_{\mathrm{sol}}\approx
4.7 $ [cf. Eq. (\ref{N}) for $\alpha =1.5$], i.e., as a matter of fact, for $%
N\geq 5$. In Fig. \ref{figure8}, we show the dependence of the effective
potential energy of the necklace clusters with vorticity $M=1$, computed as
per Eq. (\ref{pot}), on the cluster's radius for $\alpha =1$. The energy has
a local minimum at specific values of the radius, \textit{viz}., $R_{\min
}(N=5)=$ $19.8$, $R_{\min }(N=6)=$ $23.55$, $R_{\min }(N=7)=$ $27.6$, and $%
R_{\min }(N=8)=$ $31.6$, respectively. Robust propagation of the rotating
clusters at parameters corresponding to the energy minima identified in Fig. %
\ref{figure8} is confirmed by direct simulations performed in the presence
of perturbations produced by the same factor (\ref{factor}), as used above.
As shown in Fig. \ref{figure9}, the perturbed propagation is quasi-stable,
with small radial oscillations (cf. Fig. \ref{figure6}). In Fig. \ref%
{figure10}, we also display the cluster's mean radius, the mean angular
velocity, and the depth of the azimuthal modulation as functions of $\zeta $
in the course of the propagation.

\section{Conclusion}

\label{Sec V}

We have investigated ring-shaped soliton clusters (necklaces) in the
framework of the NLSE, which includes the fractional diffraction,
characterized by the respective L\'{e}vy index, $1\leq \alpha <2$, and the
competing cubic-quintic nonlinearity. Fundamental and axisymmetric vortical
solitons have their stability and instability regions in this model, the
stability of the fundamental solitons being exactly predicted by the
Vakhitov-Kolokolov criterion. Computing the effective interaction potential
for the multi-soliton clusters, the analysis predicts its minima at
particular values of the necklace's radius, at which attractive interactions
between adjacent fundamental solitons forming the cluster are balanced by
the net angular momentum. Such rotating metastable soliton clusters
withstand the action of relatively strong random perturbations, featuring
long-scale quasi-stable propagation.

These findings suggest other interesting scenarios for the generation of
complex nonlinear states in fractional dimensions. In particular, it is
relevant to extend the analysis for vector solitons in two- and
many-component settings, as well as for a $\mathcal{PT}$-symmetric system
generalizing the present one.

\section{Funding}

National Natural Science Foundation of China (11805141, 11804246); Applied
Basic Research Program of Shanxi Province (201901D211424); Scientific and
Technological Innovation Programs of Higher Education Institutions in Shanxi
(STIP) (2019L0782); \textquotedblleft\ 1331 Project\textquotedblright\ Key
Innovative Research Team of Taiyuan Normal University (I0190364); Israel
Science Foundation (1286/17).

\section{Disclosures}

The authors declare no conflicts of interest.

\end{document}